\documentclass[prd,
twocolumn,
superscriptaddress,nofootinbib,
tightenlines,showpacs,showkeys]{revtex4}
\usepackage{amsfonts,amsmath,graphicx,epsfig,amssymb}
\usepackage{mathrsfs}
\usepackage{latexsym}
\usepackage{amsxtra}
\usepackage{amsbsy}
\usepackage{amscd}
\usepackage{color}
\usepackage{bbm}
\usepackage{hyperref}
\usepackage{multirow}

\usepackage{graphics}
\usepackage{amsfonts}

\usepackage{subfigure}
\usepackage{color}
\allowdisplaybreaks

\newcommand{\be}{\begin{equation}}
\newcommand{\ee}{\end{equation}}

\def\no{\nonumber}
\def\bea{\arraycolsep .1em \begin{eqnarray}}
\def\eea{\end{eqnarray}}

\begin{document}
\title{\Large Amplitude analysis of the anomalous decay {\boldmath$\eta'\to\pi^+\pi^-\gamma$} }
\author{Ling-Yun Dai}
\email{l.dai@fz-juelich.de}
\affiliation{School of Physics and Electronics, Hunan University, Changsha 410082, China}
\affiliation{Institute for Advanced Simulation, Institut f\"ur Kernphysik
   and J\"ulich Center for Hadron Physics, Forschungszentrum J\"ulich, D-52425 J\"ulich, Germany}
\author{Xian-Wei Kang}
\email{kangxianwei1@gmail.com} \affiliation{ Institute of Physics,
Academia Sinica, Taipei, Taiwan 115}
\author{Ulf-G. Mei{\ss}ner }
\email{meissner@hiskp.uni-bonn.de}
\affiliation{Helmholtz Institut f\"ur Strahlen- und Kernphysik and Bethe Center
 for Theoretical Physics, Universit\"at Bonn, D-53115 Bonn, Germany}
\affiliation{Institute for Advanced Simulation, Institut f\"ur Kernphysik
   and J\"ulich Center for Hadron Physics, Forschungszentrum J\"ulich, D-52425 J\"ulich, Germany}
\author{Xin-Ying Song}
\email{x.song@fz-juelich.de}
\affiliation{ Institute f\"ur Kernphysik, Forschungszentrum J{\"u}lich, D-52425 J{\"u}lich, Germany}
\affiliation{ Institute of High Energy Physics, Beijing 100049, China}
\author{De-Liang~Yao}
\email{deliang.yao@ific.uv.es}
\affiliation{Instituto de F\'{\i}sica Corpuscular (centro mixto CSIC-UV), Institutos de Investigaci\'on de Paterna,
Apartado 22085, 46071, Valencia, Spain}
\begin{abstract}
In this paper we perform an amplitude analysis of
$\eta'\to\pi^+\pi^-\gamma$ and confront it with the latest BESIII
data. Based on the final-state interaction theorem, we represent the
amplitude in terms of an Omn\'es function multiplied by a form
factor that corresponds to the contributions from left-hand cuts and
right-hand cuts in the inelastic channels. We also take into account the
isospin violation effect induced by $\rho-\omega$ mixing. Our
results show that the anomaly contribution is mandatory in order to
explain the data. Its contribution to the decay width of
$\Gamma(\eta'\to\pi\pi\gamma)$ is larger than that induced by
isospin violation. Finally we extract the pole positions of the
$\rho$ and $\omega$ as well as their corresponding residues.
\end{abstract}

\maketitle

\parskip=2mm
\baselineskip=3.5mm

\section{Introduction}\label{sec:0}
There has long been interest in the study of anomalous decays, which
are driven by the chiral anomaly of QCD.\footnote{For an introduction on anomalies, see
e.g.~\cite{Bertlmann:1996xk}.} The $\eta^\prime/\eta\to\pi^+\pi^-\gamma$ decays are
typical processes for exploring the box anomaly and
investigating the $\rho-\omega$ mixing mechanism. They are useful to
extract the pion vector form factor~\cite{Hanhart2011,Kubis2015,Xiao2015,Xiao2016} and
the form factors of
$\eta^\prime/\eta\to\gamma\gamma^*$ transitions~\cite{Kubis2015,Xiao2015,Xiao2016}, helping us
to further test, e.g., the Pascalutsa-Vanderhaeghen light-by-light
sum rule ~\cite{PV1,Igor2016,DLY-MRP2017}. The so-obtained knowledge of these form factors is also
crucial in the determination of hadronic contribution to the anomalous
magnetic moment of the muon
\cite{Blum:2015you,Jegerlehner2017,DellaMorte:2017dyu}, as witnessed by
the preparation for the planned experiments at
Fermilab~\cite{Fermilab-g2} and J-PARC~\cite{J-PARC}. Furthermore,
the two processes are helpful for decoding information on the resonances
(intermediate states) such as $\rho$ and $\omega$. For instance, the
branching ratio of $\omega\to\pi\pi$ has been extracted in
Ref.~\cite{Xiao2016}.

Searching for the box anomaly in the $\eta^\prime/\eta\to\pi^+\pi^-\gamma$ decays,
is also an interesting topic on the experimental side.
For $\eta$ decay, WASA-at-COSY~\cite{WASA} and KLOE~\cite{KLOE}
have determined the relevant parameters using the approach, proposed in
Ref.~\cite{Hanhart2011}, based on chiral perturbation theory ($\chi$PT) and
dispersion theory.  For the corresponding $\eta^\prime$
decay, JADE~\cite{JADE}, CELLO~\cite{CELLO}, PLUTO~\cite{PLUTO},
TASSO~\cite{TASSO}, TPC~\cite{TPC}, and ARGUS~\cite{ARGUS} all
observed a peak shift of about +20~MeV$/c^{2}$ in the di-pion mass
spectrum, with respect to the expected position of the $\rho^{0}$. This
certainly indicates that only a contribution from the $\rho$ is not
sufficient. This issue is
also discussed in Ref.~\cite{KangBl4} for $B_{\ell4}$ decay.
In the analysis of Ref.~\cite{ZPC}, it is shown
that the contribution from the box anomaly could be essential in
$\eta^\prime\to\pi^+\pi^-\gamma$.  Later, the significance of the box
anomaly was found to be 4$\sigma$ by the Crystal Barrel (CB)
Collaboration~\cite{CB1997} with 7400 events, while the L3
Collaboration~\cite{L3-1998} claimed that the $\rho$ contribution is
sufficient to describe the data with less data ($2123\pm53$ events).
Nonetheless, the CB data is not precise enough to disentangle the effect of $\rho-\omega$ interference from others in the line shape, as shown, e.g., by Ref.~\cite{Benayoun:2009im}.
Recently, the BESIII Collaboration~\cite{Ablikim:2017fll,Fang2017} explored the
process $\eta'\to\pi^+\pi^-\gamma$ with very high statistics (of
about $9.7\times10^{5}$ events) and the $\rho-\omega$ interference
is seen for the first time in this decay. Therefore,  it is
timely to make a refined amplitude analysis of the
anomalous decay $\eta^\prime\to\pi^+\pi^-\gamma$.

In the $\eta'\to\pi^+\pi^-\gamma$ decay, the contribution of the
anomaly is significant, and hence can not be simply determined by
a tree-level amplitude from the Wess-Zumino-Witten (WZW)~\cite{WZ,Witten} term.
In Ref.~\cite{Borasoy:2004qj}, this anomalous decay is studied using $\chi$PT in
combination with a non-perturbative method based on coupled channels. Here, by
properly taking into account the effects of final-state interaction (FSI) and
$\rho-\omega$ mixing, we aim at obtaining precise information about the anomaly.
On the one hand, since the effect of three-body rescattering between the pions and
the photon is negligible due to the tiny electromagnetic interaction, a purely
strong $\pi\pi$ FSI should be sufficient. The
non-perturbative  $\pi\pi$ FSI is implemented in a
model-independent way, where the contribution corresponding to the unitary cut is
represented by an Omn\`es function. On the other hand, our treatment of
$\rho-\omega$ mixing is beyond the simplified version employed in
Ref.~\cite{Xiao2016}. In our case, the isospin-violating $\rho-\omega$
interference is constructed by invoking resonance chiral theory (R$\chi$T)
\cite{Ecker:1988te,Ecker:1989yg,RuizFemenia:2003hm,Cirigliano:2006hb,Guo2007,Dumm:2009kj,DLY2013}
(for earlier attempts using $\chi$PT with explicit vector mesons, see e.g.,
Refs.~\cite{Urech:1995ry,Kucukarslan:2006wk}). The approach we use here is a combination
of dispersion theory and chiral effective field
theory. Similar prescriptions have been widely applied to, e.g., $\gamma\gamma\to\pi\pi$
\cite{Moussallam, DLY-MRP14}, $\eta\to3\pi$ \cite{Guo:2015zqa}, $\phi/\omega\to\pi\gamma^*$
\cite{Kubis}, $D_{\ell 3}$~\cite{Yao:2017unm} and $B_{\ell 4}$ \cite{KangBl4}. With our method,
it is clearly shown that the anomaly is mandatory in order to
describe the experimental data. Besides, it is possible and also interesting
to see how much the leading-order (LO) result in the $1/N_C$ expansion of
R$\chi$T is modified by the $\pi\pi$ rescattering. For a general discussion
the role of vector mesons in anomalous processes, see e.g.~\cite{Bando:1987br,Meissner:1987ge}.

This paper is organized as follows. In Section~\ref{sec:1} we
discuss  the formalism for the $\eta^\prime\to\pi\pi\gamma$ decay amplitude. The inclusion of
the FSI effect is discussed in Section~\ref{sec:1;1}, while the isospin-violating form
factors are calculated within R$\chi$T in Section~\ref{sec:1;2}. Section~\ref{sec:1;3}
provides the final form of the decay amplitude. Section \ref{sec:2} contains our
numerical results. We fit to the experimental data and pin down all the relevant
unknown parameters in Section~\ref{sec:2;1}. In Section~\ref{sec:2;2}, we extract
the poles of the $\rho$ and $\omega$ resonances as well as their couplings to the
$\eta'\gamma$ states, which are then used to calculate the decay widths. We also
discuss the impact of isospin-violating effect on the $P$-wave phase of $\pi^+\pi^-$
scattering. Finally, we summarize and make conclusions in Section~\ref{sec:3}.
The explicit expressions of the isospin-violating form factors in R$\chi$T
are relegated to Appendix~\ref{app:isov}.

\section{ Amplitude formalism }\label{sec:1}
\subsection{Final state interaction }\label{sec:1;1}
\begin{figure}[htbp]
\vspace{-0.0cm}
\centering
\includegraphics[width=0.23\textwidth,height=0.15\textheight]{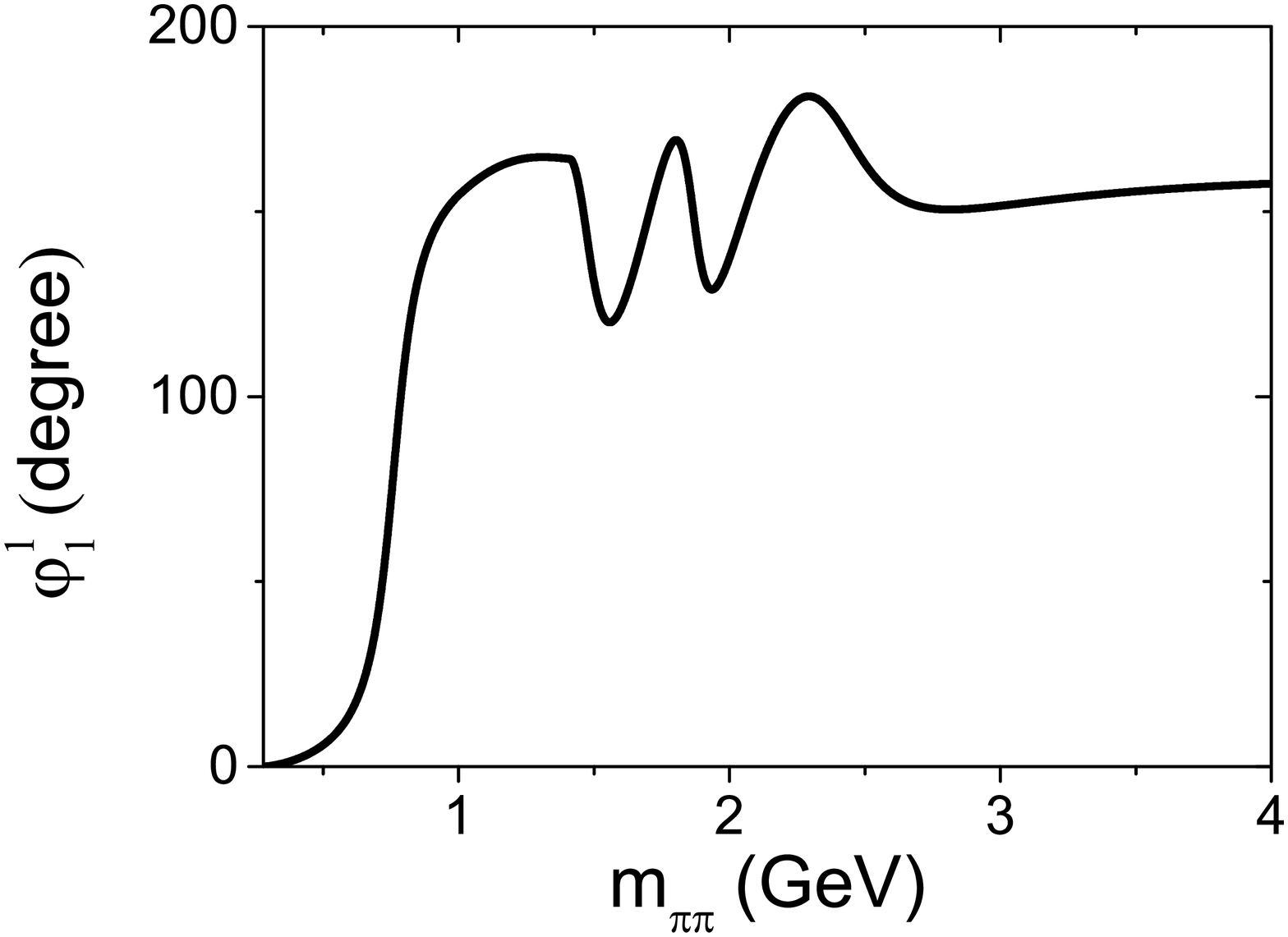}
\includegraphics[width=0.23\textwidth,height=0.15\textheight]{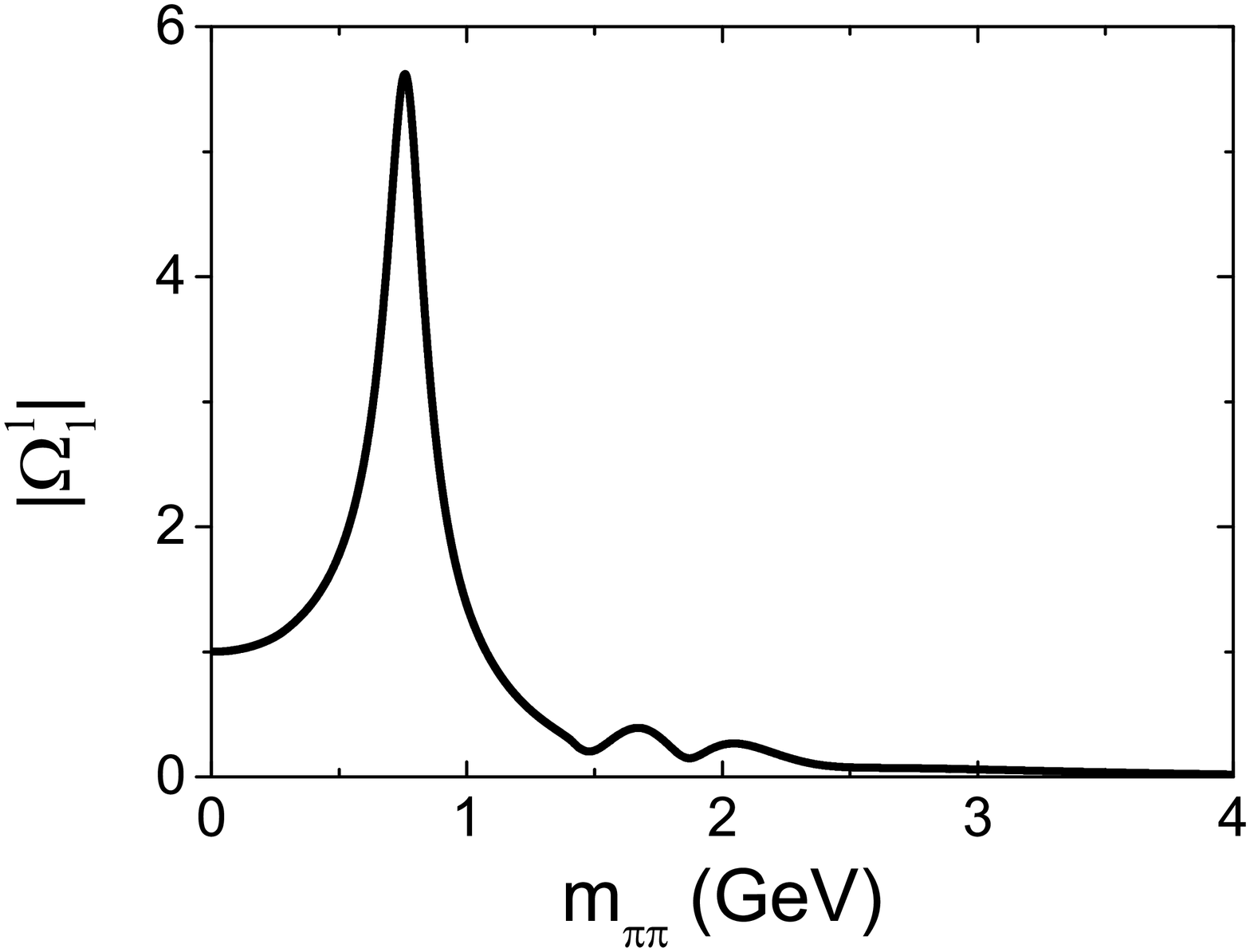}
\caption{Left: Phase of the $\pi\pi$ scattering amplitude in
$P$-wave with isospin $I=1$. Right: modulus of the
Omn\`es function. \label{fig:Omnes}}
\end{figure}

Bose and charge conjugation symmetry guarantee that
$\eta'\to\pi^0\pi^0\gamma$ is forbidden and hence we only need to
consider the mode $\eta'\to\pi^+\pi^-\gamma$.  As will be explained
below, its full amplitude is composed of partial waves
with isospin $I=1$ and odd angular momentum, i.e. $P$, $F$, $\ldots$ waves.
The isospin violation part induced mainly by the $\rho-\omega$ mixing is
of isospin $I=0$, which will be discussed in the next section.
The Lorentz-invariant decay amplitude for
$\eta^\prime(q)\to\pi^+(p_+)\pi^-(p_-)\gamma(k)$ can be written as:
\be
\mathcal{M}_\lambda=e\epsilon^{\mu\nu\alpha\beta}\epsilon_\mu(k,\lambda)
q_{\nu}p^+_\alpha p^-_\beta  \mathcal{F}_{\lambda}(s,\cos\theta) \,, \label{eq:amp;M;0}
\ee
where $\epsilon_\mu(k,\lambda)$ is the polarization of the
outgoing photon with helicity $\lambda$. The variables of the
form factor $\mathcal{F}_{\lambda}$ are chosen to be $s\equiv(q-k)^2$ and $\cos\theta$.
Here, $\theta$ is the scattering angle in the $\pi\pi$ center of mass
frame. The partial-wave decomposition of $M_{\lambda}$ reads
\bea
\mathcal{M}_{\lambda}=16\pi \sqrt{N_{\pi\pi}} \sum_{J} M_{J
\lambda}(s)d^J_{\lambda0}(\theta)(2J+1)\,,\label{eq:amp;M}
\eea
where the normalization factor should be set to $N_{\pi\pi}=2$, in
accordance with the Bose statistics of identical particles. The
isospin decomposition is given by
$M_{J\lambda}(s)=-M^1_{J\lambda}(s)/\sqrt{2}$, so that
\bea
M^1_{J\lambda}(s)&=&-\frac{e
(M_{\eta'}^2-s)\sqrt{s-4M_\pi^2}F^1_{J\lambda}(s)}{128\sqrt{2}\pi}
\,, \eea with \be F^1_{J\lambda}(s)=\int_{-1}^1 d\cos\theta
\mathcal{F}_{\lambda}(s,\cos\theta) \sin\theta
d^J_{\lambda0}(\theta)\,.\label{eq:F;int}
\ee
From the parity conservation one has $F^1_{J\lambda}(s)=(-)^J F^1_{J(-\lambda)}(s)$.
Here we only keep the lowest $P$-wave and ignore $F$- and higher
partial waves since their contributions are relatively small.
Following these constraints we have only one independent partial
wave, $M^1_{1+}(s)$ or $M^1_{1-}(s)$.  Comparing Eqs.~(\ref{eq:amp;M;0}-\ref{eq:F;int}) one soon finds $\mathcal{F}_{+}(s,\cos\theta)=-3/(2\sqrt{2}) F^1_{1+}(s)$. Notice that higher order
corrections of QED are negligible compared to hadronic FSI, thus we do not
take them into account. Finally we construct our amplitude based on Watson's FSI
theorem:
\be
\label{eq:omnesrep}
F^1_{1+}(s)=P(s)\Omega^1_1(s)\,,
\ee
with $\Omega^1_1(s)$ the so-called Omn\`es function and $P(s)$
 a polynomial. We will discuss $P(s)$ in
Section~\ref{sec:1;3}.
The Omn\`es function satisfies the following dispersion relation
\be\label{eq:Omnes}
\Omega^1_{1}(s)=\exp\left(\frac{s}{\pi} \int^\infty_{4M_\pi^2} ds'
\frac{\varphi^1_{1}(s')}{s'(s'-s)}\right) \,.
\ee
The function $\varphi^1_{1}$(s) denotes the phase of
$P$-wave elastic $\pi\pi\to\pi\pi$ amplitude,
which was given in a previous amplitude analysis of $\pi\pi$ scattering \cite{DLY2015:1,DLY2015:2}.
In Fig.~\ref{fig:Omnes}, we show the phase and modulus of the Omn\`es function.
Up to the $\overline{K}K$ threshold,
it has the same phase as that of the decay amplitude $F^1_{1+}(s)$.

\subsection{Formalism of R$\chi$T}\label{sec:1;2}
In this section, we will calculate the chiral anomaly contribution as well as the isospin-violating amplitude for
$\eta^\prime \to \pi^+\pi^-\gamma$ at LO in the large $N_C$ expansion in R$\chi$T,
with $N_C$ the number of colors.
The $\rho-\omega$ mixing is taken into account following Ref.~\cite{DLY2013}.
The relevant chiral Lagrangian can be written as~\cite{RuizFemenia:2003hm}:
\begin{equation} \label{eq:lr}
 {\cal L}_{\tiny R\chi T} = {\cal L}^{\mbox{\tiny V}}_{\mbox{\tiny kin}}
+ {\cal L}_{\mbox{\tiny int}}^{\mbox{\tiny V}}+ {\cal L}^{\mbox{\tiny GB}}_{(4)} \, .
\end{equation}
The part $ {\cal L}^{\mbox{\tiny GB}}_{(4)}$ containing
the LO operators of the chiral anomaly is~\cite{WZ,Witten}
\begin{equation} \label{eq:wzanomaly}
 {\cal L}^{\mbox{\tiny GB}}_{(4)} = i \frac{N_C \sqrt{2}}{12 \pi^2 F^3} \, \varepsilon_{\mu \nu \rho \sigma} \, \langle \partial^{\mu} \Phi \, \partial^{\nu} \Phi \, \partial^{\rho} \Phi \,  v^\sigma \rangle + ... \, ,
\end{equation}
where $F\approx 92.2$~MeV is the
pion decay constant in the chiral limit and $\Phi$ is a nonet matrix
collecting the pseudoscalar Goldstone bosons:
\bea \Phi=
 \left( {\begin{array}{*{3}c}
   {\dfrac{\pi ^0}{\sqrt{2}} +\dfrac{\eta _8}{\sqrt{6}} } & {\pi^+ } & {K^+ }  \\
   {\pi^- } & {-\dfrac{\pi ^0}{\sqrt{2}} +\dfrac{\eta _8}{\sqrt{6}}} & {K^0 }  \\
   { K^-} & {\overline{K}^0 } & {-\dfrac{2\,\eta _8}{\sqrt{6}} }  \\
\end{array}} \right)+\frac{\mathbbm{1}}{\sqrt{3}}\eta_0\ .
\eea
The mixing of the $\eta_8$ and $\eta_0$ with an angle $\theta_P$
yields the physical $\eta$ and $\eta'$ states\footnote{We
notice that more complicated schemes for $\eta-\eta'$ mixing
involving two mixing angles have been studied in e.g., Refs.~\cite{Leutwyler:1997yr,Feldmann:1998vh,Kaiser:1998ds,Guo:2015xva,ChenYH1},
which is commonly used to describe the two-photon decays well~\cite{Leutwyler:1997yr,Feldmann:1998vh,ChenYH1}. }.
\begin{equation} \label{eq:etas}
 \left( \begin{array}{c}
          \eta_8 \\ \eta_0
        \end{array}
 \right)
= \left( \begin{array}{cc}
          \cos \theta_P & \sin \theta_P \\ -\sin \theta_P & \cos \theta_P
         \end{array}
 \right) \, \left( \begin{array}{c}
                    \eta \\ \eta'
                   \end{array}
\right).
\end{equation}
The kinematic part of the vector resonances reads
\begin{equation}\label{eq:kin}
 {\cal L}^{\mbox{\tiny V}}_{\mbox{\tiny kin}} = - \frac{1}{2} \langle \nabla^{\lambda} V_{\lambda \mu} \nabla_{\nu} V^{\nu \mu} \rangle + \frac{1}{4} M_V^2 \langle
V_{\mu \nu} V^{\mu \nu} \rangle  \, ,
\end{equation}
where $M_V$ is the mass of the vector resonances in the chiral limit.
The matrix field $V^{\mu\nu}$, in antisymmetric tensor representation, incorporates the low-lying vector resonances in a nonet form
\bea
{V}^{\mu\nu}=\sum_{i=1}^8\frac{\lambda_i}{\sqrt{2}}V_i^{\mu\nu}
+\frac{\mathbbm{1}}{\sqrt{3}}V_0^{\mu\nu}\ ,
\eea
with $\lambda_i$ ($i=1,\cdots,8$) and $\mathbbm{1}$  the standard Gell-Mann matrices
and the $3\times3$ unit matrix, respectively.
The covariant derivative acting on the vector fields is defined by~\cite{Ecker:1988te}
\bea
\nabla_{\alpha} V^{\mu\nu}&=&\partial_\alpha V^{\mu\nu}+[\Gamma_\alpha,V^{\mu\nu}]\ ,\nonumber\\
\Gamma_\mu&=&\frac{1}{2}(u^\dagger(\partial_\mu-r_\mu) u+u(\partial_\mu-l_\mu) u^\dagger )\ ,\nonumber\\
l_\mu&=&v_\mu-a_\mu\ ,\quad r_\mu=v_\mu+a_\mu\ ,
\eea
where $v_\mu$ and $a_\mu$ denote external vector and axial-vector fields,
respectively. It is worth noting that the photon field $A_\mu$ can
be introduced by setting $r_\mu=l_\mu= e\,Q\, A_\mu$ with $Q={\rm
diag}\{2/3,-1/3,-1/3\}$. Furthermore,
$u=\exp\{{i\Phi}/(\sqrt{2}F)\}$.

The interaction between the vector resonances and the Goldstone bosons is described by
\be\label{eq:laggg}
{\cal L}_{\mbox{\tiny int}}^{\mbox{\tiny V}} = {\cal L}^{\mbox{\tiny V}}_{(2)}
+{\cal L}^{\mbox{\tiny V}}_{(4)}+{\cal L}^{\mbox{\tiny VV}}_{(2)}\;.
\ee
Here, the subscripts denote the chiral orders, the corresponding superscripts
imply the numbers of vector resonance. Specifically, the first term
reads \cite{Ecker:1988te}
\begin{equation} \label{eq:lv2}
{\cal L}^{\mbox{\tiny V}}_{(2)} = \frac{F_V}{2 \sqrt{2}} \, \langle V_{\mu \nu} f_+^{\mu \nu} \rangle + i \frac{G_V}{\sqrt{2}} \, \langle V_{\mu \nu} u^{\mu} u^{\nu} \rangle \, .
\end{equation}
Here, $F_V$ and $G_V$ are unknown coupling constants, that can be fixed from
certain resonance decays or from short-distance QCD constraints, see in Appendix \ref{app:isov}. The chiral building blocks, $f_+^{\mu\nu}$ and $u_\mu$
are given explicitely in Ref.~\cite{Ecker:1988te}.
The last two terms in Eq~\eqref{eq:laggg} are of odd-intrinsic parity. The pieces
relevant to our calculation can be expressed as
\cite{RuizFemenia:2003hm,Dumm:2009kj}
\begin{eqnarray} \label{eq:lv4}
 {\cal L}^{\mbox{\tiny V}}_{(4)} &=& \sum_{i=1}^7 \frac{c_i}{M_V} \, {\cal O}_{\mbox{\tiny VJP}}^i \,
+ \, \sum_{j=1}^5 \frac{g_j}{M_V} \, {\cal O}_{\mbox{\tiny VPPP}}^j \, ,\nonumber\\
 {\cal L}^{\mbox{\tiny VV}}_{(2)} &=& \sum_{k=1}^4 d_k {\cal O}_{\mbox{\tiny VVP}}^k
 \, .
\end{eqnarray}
These couplings are defined to be dimensionless. For the explicit expressions of the odd-intrinsic chiral operators
we refer the readers to Refs.~\cite{RuizFemenia:2003hm,Dumm:2009kj}.
The values of the parameters, $c_i$, $g_j$ and $d_k$, are taken from Ref.~\cite{DLY2013},
and are also given in our Appendix.~\ref{app:isov}.

The fields used in the above-mentioned chiral effective Lagrangians are
convenient for analyzing transformation properties under the chiral
group. Nonetheless, not all of them directly correspond to physical
states. In practice, the physical $\omega(782)$ and $\phi(1020)$
states are related to the octet and singlet components by a mixing
angle $\theta_V$ through
\begin{equation} \label{eq:vectos}
 \left( \begin{array}{c}
          V^8 \\ V^0
        \end{array}
 \right)
= \left( \begin{array}{cc}
          \cos \theta_V & \sin \theta_V \\ -\sin \theta_V & \cos \theta_V
         \end{array}
 \right) \, \left( \begin{array}{c}
                    \phi \\ \omega
                   \end{array}
\right).
\end{equation}
In the same manner, the $\rho-\omega$ mixing due to isospin symmetry violation
can be parameterized as
\begin{equation} \label{eq:rwmixing}
 \left( \begin{array}{c}
          \bar{\rho}^0 \\ \bar{\omega}
        \end{array}
 \right)
= \left( \begin{array}{cc}
          \cos \delta & \sin \delta \\ -\sin \delta & \cos \delta
         \end{array}
 \right) \, \left( \begin{array}{c}
                    \rho^0 \\ \omega
                   \end{array}
\right),
\end{equation}
with $\delta$ the mixing angle. The values of these mixing
angles we used are listed in Appendix \ref{app:isov}\footnote{We
notice that more complicated scheme for $\rho-\omega$ mixing with two parameters
have been studied in e.g., Refs.~\cite{Chen:2017jcw}. }.

Eventually, we are in the position to calculate the amplitudes for
$\eta^\prime\to\pi^+\pi^-\gamma$ including explicitly the WZW term as
well as the isospin violation. Throughout, we assume that the $\rho$-$\omega$
mixing is the dominant isospin breraking effect. The relevant Feynman
diagrams are displayed in Fig.~\ref{fig:isov} and the resulting amplitudes are
given in Appendix.~\ref{app:isov}.

\subsection{Isospin-violating form factor \label{sec:1;3}}
To include the dominant isospin-violating effect, we utilize the following
form for the polynomial in Eq.~\eqref{eq:omnesrep},
\bea\label{eq:DRP}
P(s)=\alpha_0+\alpha_1(s-4M_\pi^2)+P^{i.v.}(s)\, .
\eea
The contributions from left hand cuts (l.h.c) and inelastic right hand cuts (r.h.c) are
ascribed  to the polynomial in front,
and $P^{i.v.}(s)$ stands for the contribution from isospin violation.
Following the experimental paper~\cite{Ablikim:2017fll}, we set
\bea
P^{i.v.}(s)&=&\beta_0 F^{i.v.}_{tree}(s)\;,\;\no\\
F^{i.v.}_{1+}(s)&=&P^{i.v.}(s)\Omega^1_1(s)\;,\label{eq:FIV}
\eea
with $F^{i.v.}_{tree}(s)$ the isospin-violating form factor given in Eq.~(\ref{eq:A;Fiv}).  The phase of $\pi\pi$ rescattering is included by the Omn\`es function, and the l.h.c
and inelastic r.h.c contributions of isospin violation are absorbed in the parameter
$\beta_0$. In principle there should be more terms rather than a single $\beta_0$,
but in practice we find that one parameter is good enough to describe the data well.

We notice that  $F^{i.v.}_{tree}(s)$ has a sizeable imaginary part around
$\sqrt{s}=M_\omega$. It modifies the $\pi^+\pi^-$ phase in the vicinity of
$\sqrt{s}=M_\omega$. Nevertheless, this
is helpful for us to obtain $\rho-\omega$ mixing
exactly in the $\pi^+\pi^-$ FSI. Previous dispersive analyses
\cite{Colangelo01,KPY} and also experiments
\cite{CERN-Munich,Protopopescu1978,Estabrooks1974} take only the
contribution of $\rho$ in their partial wave analysis, thus the
information of isospin-violating part, i.e. the contribution from the $\omega$,
is lacking in this region. This will be discussed with more details in the next section.

\begin{figure}[htbp]
\vspace{-0.0cm}
\centering
\includegraphics[width=0.48\textwidth,height=0.25\textheight]{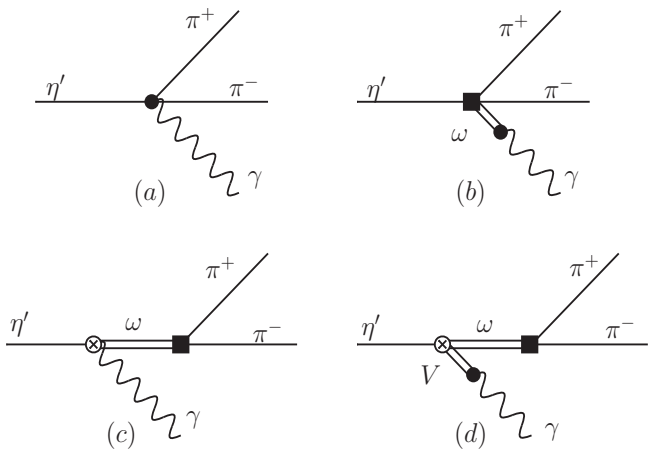}
\caption{ The Feynmann diagrams of the Wess-Zumino-Witten (WZW) term (Fig.a),
and the isospin violation amplitudes (Figs.b-d).
The symbol \lq V' represents the vector resonances $\rho$, $\omega$, and $\phi$. The odd-intrinsic parity and isospin-violating vertices are denoted by the circled crosses and black squares, respectively. Note that in Fig (b) the black square also represents an odd-intrinsic parity vertex. \label{fig:isov}}
\end{figure}

For the isospin-violating form factor $F^{i.v.}_{tree}(s)$, we only need
to calculate the diagrams, (b), (c) and (d), in Fig.~\ref{fig:isov}.
The results are given in Appendix \ref{app:isov}.
Finally, the total  amplitude is given by:
\bea
F^1_{1+}(s)=[\alpha_0+\alpha_1(s-4M_\pi^2)+\beta_0 F^{i.v.}_{tree}(s)]\Omega^1_1(s)\,. \label{eq:F}
\eea
Note that $F^1_{1+}(s)$ contains an isospin-violating part ($I=0$), though its superscript
is labeled by \lq1', corresponding to $I=1$. With this amplitude one can get the
formula for the di-pion mass spectrum:
\bea
\frac{d\Gamma}{d\sqrt{s}}=\frac{3\alpha s^{3/2} \rho(s)^3 (M_{\eta'}^2-s)^3 |F^1_{1+}(s)|^2}{2048\pi^2M_{\eta'}^3}\;, \label{eq:dGds}
\eea
with $\rho(s)=\sqrt{1-4M_\pi^2/s}$. When fitting to the invariant mass spectrum of
BESIII \cite{Ablikim:2017fll}, we need to multiply it by a normalization factor $N$.
In Ref.~\cite{Kubis2015}, the anomaly is obtained at the point where $s=t=u=0$, in the chiral limit.
Similarly, we define the anomaly as $-e\mathcal{F}_+(0,\cos\theta)$. Notice that here $\theta$ is a dummy variable in $\mathcal{F}_+(0,\cos\theta)$. We have
\be\label{eq:anomaly}
A=\frac{3e}{2\sqrt{2}}(\alpha_0-4M_\pi^2\alpha_1+\beta_0F^{i.v.}_{tree}(0))\,,
\ee
where $F^{i.v.}_{tree}(0)=0.155$~GeV$^{-3}$, see in Appendix \ref{app:isov}.

\section{ Numerical results  }\label{sec:2}

\subsection{Fit to experimental data }\label{sec:2;1}
In this section, we fit to the invariant di-pion mass spectrum by BESIII~\cite{Ablikim:2017fll}.
The decay width of $\eta'\to\pi^+\pi^-\gamma$ provided by the Particle Data Group (PDG)~\cite{PDG2016} is
implemented in our fit to constrain the unknown normalization factor $N$.
The following fits are performed:
\begin{itemize}
\item[1)] {\bf Fit~1}: We ignore isospin violation ($\beta_0=0$). The best-fit parameters
are collected in the second column of Table~\ref{tab:para}.
\item[2)] {\bf Fit~2}: We include the contribution of isospin violation and $\alpha_0$
is fixed by Eq.~(\ref{eq:alpha0}). The results are shown in the third column in
Table~\ref{tab:para}.
\item[3)] {\bf Fit~3}: As in Fit~2 we include the isospin violation, but set $\alpha_0$
to be a free parameter.
The results are shown in the fourth column in Table~\ref{tab:para}.
\end{itemize}
In fact, a chiral matching can be imposed to fix $\alpha_0$~\cite{DLY2012}. At low energies,
our amplitude is required to coincide with the one calculated from the
LO WZW term~\cite{WZ,Witten}.  The corresponding Feynman diagram is shown in
Fig.~\ref{fig:isov} (a) and its form factor is given in Eq.~(\ref{eq:WZW}).
We choose $s=4M_\pi^2$ as the matching point to avoid any complication caused
by the $\alpha_1$ term in Eq.~(\ref{eq:DRP}).  After the chiral matching described
above, one obtains
\bea\label{eq:alpha0}
\alpha_0 = \frac{\sqrt{2}N_C}{18 \sqrt{3}\pi^2 F^3\Omega^1_1(4M_\pi^2)}(\sin\theta_P+\sqrt{2}\cos\theta_P)\,.
\eea
The value of $\alpha_0$ is fixed to be $14.37$~GeV$^{-3}$ provided $\theta_P=-21.37^\circ$. Note that $\Omega^1_1(4M_\pi^2)=1.159$ is a real number.
This corresponds to Fit~2. In contrast, if we use double-angles-mixing scheme (DAMS), we obtain $\alpha_0=15.17$~GeV$^{-3}$, where the angles and decay constants are taken from \lq NNLO Fit-A' of Ref.~\cite{Guo:2015xva}. In this case $\alpha_0$ is fairly increased and still faraway from that of Fit~3. In the absence of the high statics data of $\eta\to\pi^+\pi^-\gamma$, one can not reach a definite conclusion on how DAMS will improve the calculation.

\begin{figure}[!phtb]
\vspace{-0.0cm}
\includegraphics[width=0.48\textwidth]{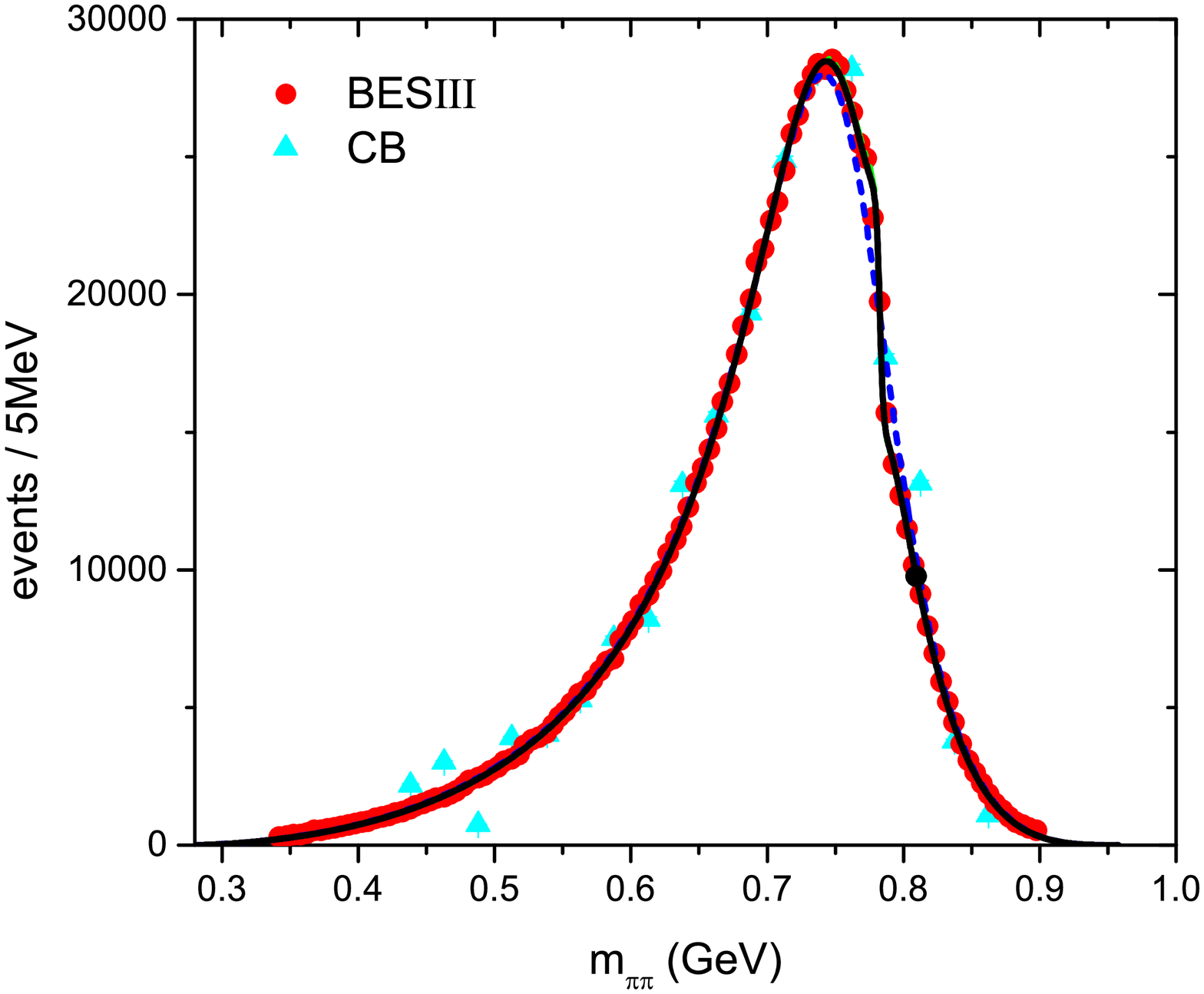}
\includegraphics[width=0.48\textwidth]{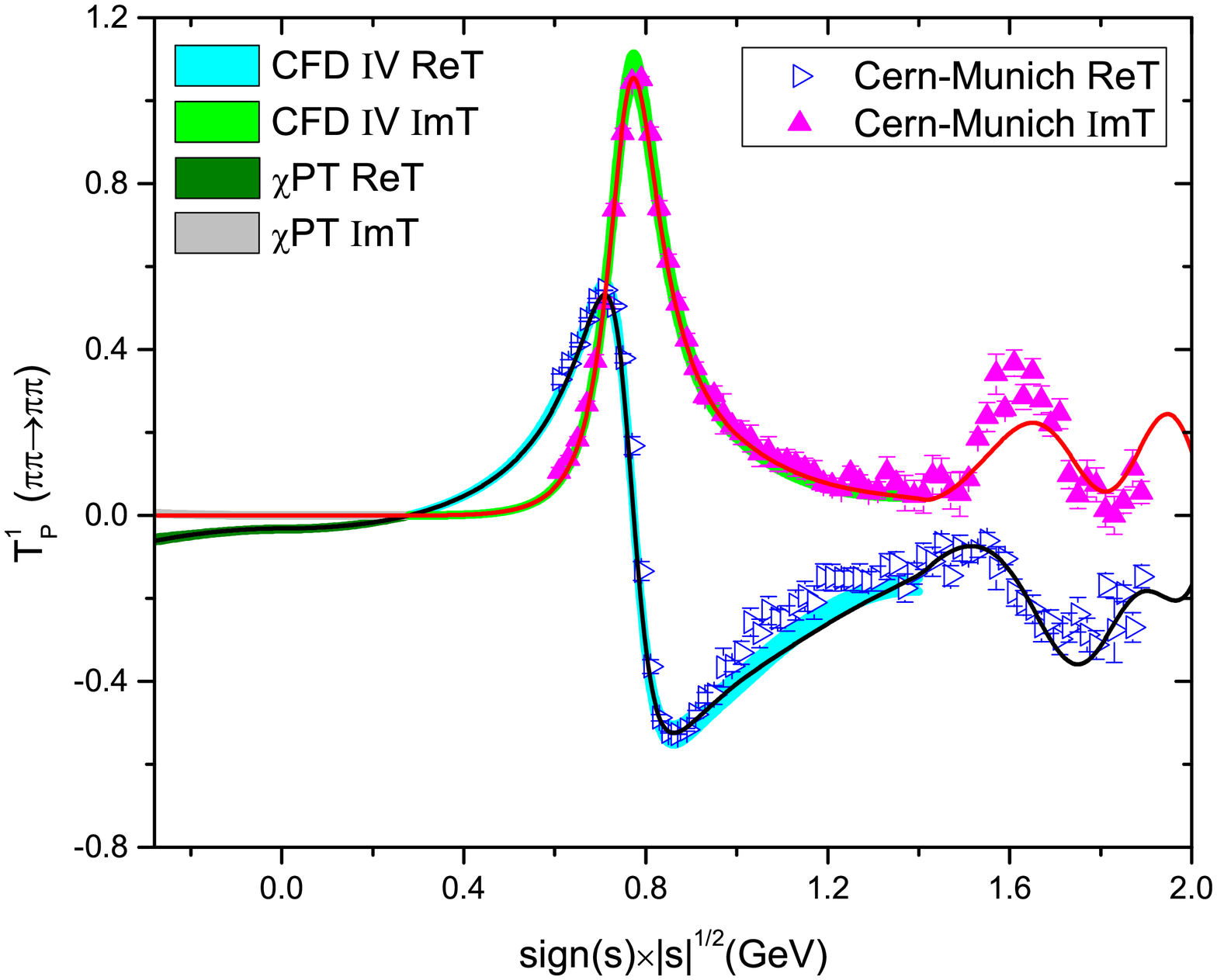}
\caption{ The top figure corresponds to the fit to the invariant mass spectrum of
$\eta'\to\pi\pi\gamma$.
The black solid line is the one of Fit~3, the blue dotted line is of Fit~1.
As explained in the text, Fit~2 is almost indistinguishable from  of Fit~3
and thus not shown.
The BESIII data is from \cite{Ablikim:2017fll} and Crystal Ball data from\cite{CB1997}.
The bottom figure corresponds to the fit to the experiment data of $\pi\pi\to\pi\pi$ $P$-wave.
The Cern-Munich data is from \cite{CERN-Munich}, and the olive and light grey bands in the low energy region are from Refs.~\cite{Gasser1985,Bijnens1994,Pelaez2002,DLY11}.
\label{fig:fit}}
\end{figure}

Actually, one may treat $\theta_P$ as a free parameter, while $\alpha_0$ is always fixed
by using Eq.~(\ref{eq:alpha0}) during the fit procedure.
A good fit can be obtained with $\theta_P\approx-10.9\pm0.5^\circ$, which is compatible
with the determinations in Refs.~\cite{Kubis2015,KangDs,Ambrosino:2006gk,Hietala}.
However, the resulting decay widths involving $\eta'$, $\Gamma_{\eta'\to\rho\gamma}$,
$\Gamma_{\eta'\to\omega\gamma}$ and $\Gamma_{\phi\to\eta'\gamma}$, are now deviated about 30-60 percents from the previous determinations in Ref. \cite{DLY2013}.
Thus, we fix $\theta_P$ at $-21.37^\circ$~\cite{DLY2013}  and set $\alpha_0$ free.
This is Fit~3. Note that here Eq.~\eqref{eq:alpha0} is not implemented as a
constraint. In Fit~3 the BESIII data can be well described and, furthermore,
the previous results in Ref.~\cite{DLY2013} for the above-mentioned decay
widths are untouched.
For comparison, the invariant mass spectrum, based on the fitted
values of the parameters from Fit~1, Fit~2 and Fit~3, are shown simultaneously in
Fig.~\ref{fig:fit}. Note that the Crystal Ball data points are superimposed on the plot.
\begin{table}[h!]
{\footnotesize
\begin{center}
\begin{tabular}{|c||c|c|c|c|c|c|}
\hline
                           & Fit 1            & Fit 2           & Fit 3          \\
\hline\hline
$\alpha_0$(GeV$^{-3}$)     &  17.91$\pm$0.23  & 14.37           & 18.41$\pm$0.19             \\
$\alpha_1$(GeV$^{-5}$)     &  11.78$\pm$0.18  & 10.29$\pm$0.12  & 12.37$\pm$0.17     \\
$\beta_0$                  &     -            & 0.132$\pm$0.002 &  0.150$\pm$0.002                  \\
N ($\times10^8$)           &  0.852$\pm$0.021 & 1.25$\pm$0.07   &  0.788$\pm$0.016   \\
$\chi ^2_{{\rm average}}$                 &  12.3            & 2.66             &  1.74              \\
$\Gamma_{\eta'\to\gamma\pi^+\pi^-}$(keV) & 57.3$\pm$8.6   & 38.9$\pm$8.2  &  62.0$\pm$5.9    \\
anomaly (GeV$^{-3}$)       & 5.46$\pm$0.42    & 4.36$\pm$0.50  &  5.61$\pm$0.29      \\
\hline
\end{tabular}
\caption{\label{tab:para} Results for the different fits explained in the text. The
$\chi ^2_{{\rm average}}$ is the total $\chi^2$ of the  invariant mass spectrum~\cite{Ablikim:2017fll}
divided by the number of data points. The PDG~\cite{PDG2016} value of
$\Gamma(\eta'\to\gamma\pi^+\pi^-)$ is $57.3\pm1.0$~keV. The uncertainty is given from the fit. }
\end{center}
}
\end{table}

For Fit~1, the invariant mass spectrum is represented by the blue dotted line in
Fig.~\ref{fig:fit}. Obviously, there is no $\rho$-$\omega$ mixing structure appearing
in the energy region $\sqrt{s}\in[0.76, 0.8]$~GeV. This is not surprising as we do not
take the isospin violation into account. The resulting decay width of
$\eta'\to\pi^+\pi^-\gamma$ is $57.3\pm8.6$~keV is absolutely in agreement with the
value given by PDG.
For Fit~2, the fit to the BESIII data is much better. The invariant mass spectrum of
Fit~2 is almost indistinguishable from that of Fit~3,
thus we do not show it in Fig.~\ref{fig:fit}. However, the decay width of
$\eta'\to\pi^+\pi^-\gamma$, which is $38.9\pm8.2$~keV now,
deviates from the value of PDG \cite{PDG2016} by 32\%. The anomaly is
estimated to be $4.36$~GeV$^{-3}$.
For Fit~3, the fit is rather good.
The fit quality is improved both in the low energy region and the
$\rho-\omega$ mixing region, compared to Fit~2.
This is reasonable as, by tuning the free parameter $\alpha_0$,  the amplitude at
low energies can be adjusted, and the same holds for the
isospin violation part in the high-energy region.
We notice that $\alpha_0$ is shifted by 28\% compared to that of Fit~2, which also
quantifies how much the anomaly can affect the fit.
Consequently, the anomaly is  $5.61\pm0.29$~GeV$^{-3}$, shifted by 29\%  from the
one given by matching to tree-level amplitude of the WZW term.
Since the tree-level amplitude is calculated in the large $N_C$ expansion, a typical $1/N_C$ correction is reasonable.
Considering also the improved fit quality of Fit~3 with respect to Fit~2, the
correction of order $1/N_C$ to the tree-level amplitude of WZW term in anomalous
decay process, is not only reasonable but also necessary.
The normalization factor in Fit~3 is similar to the one in Fit~1 but decreases a
lot compared to that of Fit~2.
The reason is that, when $\alpha_0$ and $\alpha_1$ increase, the normalization
factor has to decrease so as to compensate for the amplitude $F^1_{1+}(s)$.
Comparing the quality of these fits, we consider Fit~3 to be the reference result.

\begin{figure}[!tbph]
\vspace{-0.0cm}
\includegraphics[width=0.48\textwidth,height=0.28\textheight]{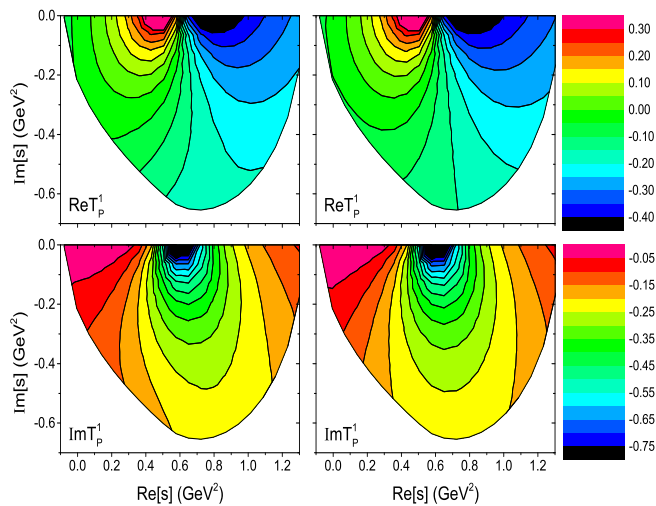}
\caption{ Comparison of our amplitudes with the ones from the Roy equation
analysis in the
domain where the Roy equations work. On the left side are real and
imaginary parts of our amplitudes, and on the right side are those from
Roy equation~\cite{Colangelo01}. \label{fig:fit;Roy}}
\end{figure}

In the next section, in order to extract the couplings of the resonances,  the $P$-wave
$\pi\pi$ scattering amplitude is needed in addition to the decay amplitude of
$\eta^\prime\to\pi^+\pi^-\gamma$ described above. To get the $P$-wave $\pi\pi$ scattering
amplitude with isospin $I=1$, we adopt the following representation
\be
T^1_P(s)=(s-4M_\pi^2)\Omega^1_1(s)\sum_{i=0}^{6} c_i(s-4M_\pi^2)^{i}\;,\label{eq:T}
\ee
with the $c_i$ unknown constants. The two constants, $c_0$ and $c_1$, are fixed by
the relevant threshold parameters: scattering length $\sim0.0387\pm0.0012~M_\pi^{-3}$
and slope parameter $\sim0.0051\pm0.0026~M_\pi^{-5}$ \cite{KPY}. The Cern-Munich
data~\cite{CERN-Munich}, CFDIV amplitude \cite{KPY}, and the amplitude from the Roy
equation analysis~\cite{Colangelo01}\footnote{We replace the phase in the energy region of $0.8-1.4$~GeV
by more recent analysis \cite{KPY,DLY-MRP14}. The results are in agreement with that
of the original paper \cite{Colangelo01}.} on the complex $s$-plane are fitted to
pin down the other constants. Finally we have all the values for the parameters,
collected in Table~\ref{tab:para;T}.
\begin{table}[hbtp]
\vspace{-0.0cm}
\begin{center}
\tabcolsep=0.0cm
\begin{tabular}{l l l l}\hline\hline
 $c_{0}$=0.4283  &~~~~$c_{1}$=$-$0.2959     & ~~ $c_{2}$=0.6173(16)~~ &~~  $c_{3}$=$-$0.7092(11) \\
 $c_{4}$=0.3774(4)~~    &~~$c_{5}$=$-$0.0909(1)~~   &~~   $c_{6}$=0.0081(1) & \\
 \hline\hline
 \end{tabular}
\caption{\label{tab:para;T}The parameters of the $\pi\pi\to\pi\pi$ P-wave, given in Eq.~(\ref{eq:T}). The uncertainty is from MINUIT. The unit of $c_j$ is GeV$^{-2j}$. }
\end{center}
\end{table}
The resulting $T^1_P(s)$ amplitude on the real axis is shown in the bottom panel of
Fig.~\ref{fig:fit}. Its analytic continuation to the complex $s$-plane, confronted with
that of the Roy equation analysis \cite{Colangelo01}, is shown in Fig.~\ref{fig:fit;Roy}.
We do not plot the amplitude on the upper half of $s$-plane,
as it is readily obtainable from the one on the lower half of $s$-plane according to
the Schwarz reflection principle.
The contribution of l.h.c to the shade region of $s$-plane, as shown in Fig.~\ref{fig:fit;Roy},
is properly implemented by fitting to the data as well as the amplitude of Roy equation in the presence of crossing symmetry.
We see that our amplitudes are quite similar to the ones from the Roy equation
analysis on the complex $s$-plane. The distribution of contours is in good
agreement with each other. Moreover, their gradient variations, the shading of
the color from blue to red, are compatible. Nevertheless, amplitudes on the
edge of the domain, shown in Fig.~\ref{fig:fit;Roy}, are less consistent with a difference $\leq0.1$. This means our $T^1_P(s)$ amplitude
is constrained rather well on the complex plane, allowing for a reliable  extraction of
the poles and residues.

\subsection{Couplings of {\boldmath$g_{\eta'V\gamma}$}}\label{sec:2;2}
With a specific amplitude, the couplings of a resonance are
defined by the residues of the pole on the complex $s$-plane.  Based on the results
of Fit~3, the absolute values of the couplings $g_{\eta'V\gamma}$, $V\in\{\rho,\omega\}$,
extracted on the appropriate Riemann sheets are compiled in Table~\ref{tab:pole}.
\begin{table*}[phtb!]
{\footnotesize
\begin{center}
\begin{tabular}{|c|c|c|c|c|c|c|}
\hline
\multirow{2}{*}{\rule[-0.5cm]{0cm}{1cm}State}      &   pole location      & $|g_{V\pi\pi}|$       & $|g_{\eta' V\gamma}|$     & ~$\Gamma(V\to\pi\pi)$~     &~$\Gamma(\eta'\to X\gamma)$~   &~$\Gamma(\eta'\to \pi^+\pi^-\gamma)$~                \\
\multirow{2}{*}{}  &    MeV    & MeV  &  MeV & MeV &  keV  &   keV  \\
\hline\hline
$\rho$           &762.7(23)$-$i 68.3(55) & 340.1(60)   & 20.1(9)    &   141.2(53)   & 56.6(53)  &  56.6(53)     \\
$\omega$         &  782.56(12) $-$ i4.24(4)  & 10.4(4)     & 5.68(74)   &    0.130(1)   &  4.10(97) &  0.0675(160)   \\
A(anomaly)       & -                 &  -                & -          &  -            &  -        &  3.34(35)          \\
total            & -                 &  -                & -          &  -            &  -        &  62.0(59)          \\
\hline
\end{tabular}
\caption{\label{tab:pole} Predictions based on Fit~3, as explained in the text.
The following PDG~\cite{PDG2016} values are used:  $\Gamma(\eta'\to\gamma\pi^+\pi^-)
= 57.3\pm1.0$~keV and $\Gamma(\eta'\to\omega\gamma)=5.16\pm0.26$~keV. }
\end{center}
}
\end{table*}

The definition of the couplings on the appropriate Riemann sheet is given as
\be
{{M}^1_{1+}}^{\rm II}(s)=\frac{e~g_{\eta'V\gamma}~g_{V\pi\pi}}{s_{V}-s}\;,\;\;\;\;
{T^1_P}^{\rm II}(s)=\frac{g_{V\pi\pi}^2}{s_{V}-s}\;,
\ee
where \lq II' denotes the second Riemann sheet.  ${M}^1_{1+}$ is the
$\eta'\to\pi\pi\gamma$ decay amplitude given in Eq.~(\ref{eq:amp;M}) and
$T^1_P(s)$ is the $P$-wave $\pi\pi\to\pi\pi$ scattering amplitude from Eq.~(\ref{eq:T}).
What those couplings mean in terms of decay width is provided as
\bea
\Gamma(\eta'\to V\gamma)&=& \frac {16\pi\alpha (M_{\eta'}^2-M_V^2)
|g_{\eta'V\gamma}|^2}{ M_{\eta'}^3}\;, \label{eq:Getap}\\
\Gamma(V\to \pi\pi)&=& \frac {\rho(M_V^2)|g_{V\pi\pi}|^2}{ M_V}\;,\label{eq:Gpipi}
\eea
with $\alpha$ the usual QED fine structure constant.

To obtain the couplings of $\omega$, 
we adopt the standard Blatt-Weisskopf barrier factor representation~\cite{DLY-MRP14}
\be\label{eq:cVpipi}
g_{\omega\pi\pi}^2(s)=\frac{M_{\omega}~\Gamma_{\omega}~\text{BR}_{\omega\to\pi\pi}
~\mathcal{Q}(M_{\omega}^2)}{\rho(M_{\omega}^2)~\mathcal{Q}(s)} \;,
\ee
with $\mathcal{Q}(s)=1+q^2/(s-4M_\pi^2)$ and $q$ is chosen to be $1$~GeV. The mass,
width and $\pi\pi$-mode branching ratio of the $\omega$ are taken from PDG~\cite{PDG2016}:
$M_{\omega} =782.65\pm0.12$~MeV,  $\Gamma_{\omega} = 8.49\pm0.08$~MeV and
$\text{BR}_{\omega\to\pi\pi}=0.0153\pm0.0012$.
Notice that the width is rather small and thus we can ignore its energy dependence\footnote{In a more dedicated way,
one can use the standard Breit-Wigner formalism to represent the $\pi\pi\to\omega\to \pi\pi$ amplitude\cite{DLY-MRP14}:
\be\label{eq:T;omega}
{T^1_{P}}_{\omega}=\frac{g_2(s)^2}{ M^2-s-i \rho_1(s)g_1^2(s)-i \rho_2(s)g_2^2(s)-i \rho_3(s)g_3^2(s) }.\nonumber
\ee
Here 1, 2, 3 represents the $\pi\gamma$, $\pi\pi$ and $\pi\pi\pi$ channels, respectively.
Following it one can extract out the pole and residue on the (-,-,-) plane.
As we have checked, the results obtained in this way are quite the same as what we
obtained in Table~\ref{tab:pole}.
}.
To proceed, we define the coupling $g_{\eta'\omega\gamma}$ through
\bea
g_{\eta'\omega\gamma}=\frac{  \beta_0  (M_{\eta'}^2-s_\omega)\sqrt{3s_\omega}\rho(s_\omega)
(F_c+F_d)_{s_\omega}\Omega^1_1(s_\omega)}{256\pi~{\rm BW}[\omega,s_\omega]
~g_{\omega\pi\pi}(s_\omega)},\no\\
\label{eq:getag}
\eea
where $s_\omega=M_\omega^2-i M_\omega \Gamma_\omega$ and ${\rm BW}[\omega,s]$ is the Breit-Wigner
representation given in Eq.~(\ref{eq:BW}).

To get the anomaly contribution to the width of $\Gamma(\eta'\to\pi^+\pi^-\gamma)$, for
simplicity we use $F^1_{1+}(0)$ instead of $F^1_{1+}(s)$, when integrating over $\sqrt{s}$
in Eq.~(\ref{eq:dGds}). The differential decay width of the mode $\eta'\to
\omega\gamma\to\pi^+\pi^-\gamma$ is written as
\bea
\frac{d\Gamma}{d\sqrt{s}}=\frac{3\alpha s^{3/2} \rho(s)^3 (M_{\eta'}^2-s)^3
|\beta_0(F_c+F_d)\Omega^1_1(s)|^2}{2048\pi^2M_{\eta'}^3}\; .\no\\
\label{eq:dGds;omega}
\eea
The contribution through the intermediate $\rho$ meson is the same as
$\Gamma(\eta'\to\rho\gamma)$ since $\rho$ decays into $\pi\pi$ to hundred percent,
see also Eq.~(\ref{eq:G;cascad}).
Finally, we obtain the poles and couplings, based on Fit~3, and compute the decay
widths with the help of Eqs.~(\ref{eq:Getap}), (\ref{eq:Gpipi}) and (\ref{eq:dGds;omega}).
The results  are shown in Table~\ref{tab:pole}.
Our uncertainty is from the fit, combined the error from MINUIT and the systematic one: the correlation between the coefficients, see Eqs.~(\ref{eq:F},\ref{eq:T}), and the uncertainty of the phase in the Omn\`es function.
Our estimation shows that the systematic error dominates the uncertainty.

We notice that the pole position of $\rho$ is compatible with the one obtained by Pad\'e approximants in \cite{Masjuan:2014psa}.
The $\rho$ pole position is shifted a bit compared to the Breit-Wigner mass and width given by PDG \cite{PDG2016},
while this is not the case for $\omega$. The reason is that $\rho$ is much wider and the pole is farther away from the real axis, thus \lq narrow resonance approximation' is not good enough to describe the amplitude.
With the pole locations one can extract out the residues and thus determine the contribution to $\Gamma(\eta'\to \pi^+\pi^-\gamma)$ from each resonance, separately.

We obtain $\Gamma(\eta'\to\omega\gamma\to \pi^+\pi^-\gamma)=67.5\pm16.0$~eV.
Since the $\omega$ is rather narrow,  we can use the sequential decay formula
\cite{Kang3872,ChengHY2010} to do a cross check. The decay width in this way is given by
\bea
\Gamma(\eta'\to\omega\gamma\to\pi\pi\gamma)
=\frac{\Gamma(\eta'\to\omega\gamma)\Gamma(\omega\to\pi\pi)}{\Gamma_\omega}\,.
\label{eq:G;cascad}
\eea
In combination with Eqs.~(\ref{eq:Gpipi}) and (\ref{eq:cVpipi}), we obtain
$\Gamma(\eta'\to\omega\gamma\to\pi^+\pi^-\gamma)=62.8\pm19.8$~eV, which is in a good
agreement with the one calculated from Eq.~(\ref{eq:dGds;omega}).
With  $\Gamma(\eta'\to\omega\gamma\to \pi^+\pi^-\gamma)=67.5\pm16.0$~eV and
$\Gamma(\omega\to \pi\pi)=0.130\pm0.010$~MeV taken from PDG, we obtain
$\Gamma(\eta'\to\omega\gamma)=4.41\pm1.04$~keV. This is rather close to the
value of $4.10\pm0.97$~keV,
obtained in the way described by Eqs.~(\ref{eq:Getap}) and (\ref{eq:getag}).
Comparing these two decay widths, we notice that the one given by Eqs.~(\ref{eq:dGds;omega},\ref{eq:G;cascad}) contains the more dedicated energy dependence and is closer to that of PDG, thus we adopt $\Gamma(\eta'\to\omega\gamma)=4.41\pm1.04$~keV as the optimal one.

It should be pointed out that $g_{\eta'\omega\gamma}$ is correlated to $g_{\omega\pi\pi}(s_\omega)$,
as can be seen from Eq.~(\ref{eq:getag}). Such a correlation is propagated to the decay
widths, see Eq.~(\ref{eq:G;cascad}). With $\Gamma(\eta'\to\omega\gamma\to \pi^+\pi^-\gamma)
=67.5\pm16.0$~eV, if we fix $\Gamma(\eta'\to \omega\gamma)=5.16\pm0.26$~keV instead
of $\Gamma(\omega\to\pi\pi)$, we get $\Gamma(\omega\to \pi\pi)=0.111\pm0.026$~MeV
or $\text{BR}_{\omega\to\pi\pi}=1.31\pm0.31\%$ from Eq.~(\ref{eq:G;cascad}).
This is compatible with the previous analysis
given by Ref.~\cite{Xiao2016}, shown in Table~2 therein.

The contributions to the decay width of $\Gamma(\eta'\to \pi^+\pi^-\gamma)$, induced by
intermediate $\rho$ and $\omega$ as well as anomaly \lq A', are given as follows:
$56.6\pm5.3$~keV through $\rho$, $67.5\pm16.0$~eV through $\omega$, and
$3.34\pm 0.35$~keV through the anomaly.  The contact isospin-violating term,
shown in Fig.~\ref{fig:isov} (b), contributes to the anomaly part too. However, its
contribution is tiny and hence neglected. It is found that the $\rho$ dominates the
contribution and the anomaly contributes more than the $\omega$. This is not surprising
as the $\rho$ dominates the $P$-wave and the anomaly contributes as a background in the
whole energy region, while the $\omega$ only acts in a small region and rarely decays into $\pi\pi$.

For the decay widths of $\eta'\to V\gamma$, we obtain $\Gamma(\eta'\to \rho\gamma)
= 56.6\pm5.3$~keV, while in LO R$\chi$T \cite{DLY2013} it is $53.7$~keV. Likewise,
$\Gamma(\eta'\to \omega\gamma)=4.41\pm1.04$~keV (or $4.10\pm0.97$ through residue in Eqs.~(\ref{eq:Getap},\ref{eq:getag})), while in LO R$\chi$T it is $5.12$~keV. Since in LO R$\chi$T
the uncertainty is roughly $1/3$ when truncating the large $N_C$ expansion, the widths
obtained here and in LO R$\chi$T are compatible if the errors are taken into account.
This shows us how much the $\pi\pi$ FSI affects the strong interaction between the
lightest vector and pseudoscalar mesons. Compared to the predictions given by R$\chi$T,
the FSI effect is sizable but still within the uncertainty.

\begin{figure}[!phtb]
\vspace{-0.0cm}
\centering
\includegraphics[width=0.48\textwidth]{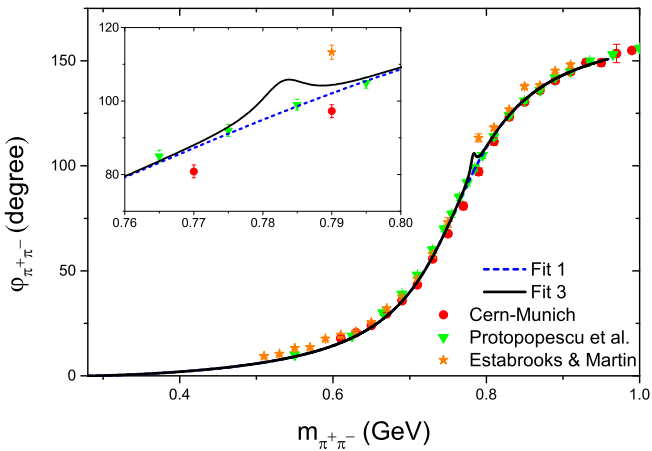}
\caption{Phase of $\pi^+\pi^-$ $P$-wave from our amplitude analysis. The blue dashed
line is from Fit~1 and the black solid line from Fit~3.
The data sets of are from Ref.~\cite{CERN-Munich} (red circles),
 from Ref.~\cite{Protopopescu1978} (green triangles)
and  from Ref.~\cite{Estabrooks1974} (orange stars), respectively.
\label{fig:phase;w}}
\end{figure}

In addition, we find that in the whole kinematical region the phase of $\pi^+\pi^-$
based on Fit~3 is in good agreement with the one based on Fit~1, i.e., the
phase of $T^1_P(s)$, except for the energy region around $\sqrt{s}=M_\omega$,
see Fig.~\ref{fig:phase;w}.
Obviously, there is a bump around the $\omega$ mass when taking into account
the $\rho-\omega$ mixing.
Note that there is no such bump for $\pi^{\pm}\pi^0$, as the $\omega$ is an isospin singlet.
It should be noted that for $\eta'\to\pi^+\pi^-\gamma$ the l.h.c (mainly from the $a_2$)
and inelastic r.h.c (above $\overline{K}K$ threshold) are too far away from the energy
region we are working. Thus the polynomial $P(s)$, in which the l.h.c and r.h.c are absorbed,
should be smooth in the physical energy region.
This suggests that, to include isospin violation from the $\rho-\omega$ mixing
in the line shape of $\eta'\to\pi^+\pi^-\gamma$,
it is more convenient to change the phase of $\pi^+\pi^-$ in the Omn\'es function than
to include a complicated complex form factor in $P(s)$ function.
We notice that there is only one data point from \cite{Protopopescu1978} located at
around $\sqrt{s}=M_{\omega}$, but it was obtained with low resolution. Most
experiments only take into account the isospin-conserved part (the contribution from
the $\rho$)~\cite{CERN-Munich,Protopopescu1978,Estabrooks1974}.
To reach a more definite conclusion on the phase caused by isospin violation,
more accurate experimental measurement in the $\sqrt{s}=M_{\omega}$ region and more
careful theory analysis are required. Here we just point out that, through our amplitude
analysis of the $\eta'\to\pi^+\pi^-\gamma$ data provided by Ref.~\cite{Ablikim:2017fll},
there should be a bump of the $\pi^+\pi^-$ $P$-wave phase around the energy
point $\sqrt{s}=M_{\omega}$.

\section{ Conclusion }\label{sec:3}
In this paper we have studied the process of $\eta'\to\pi^+\pi^-\gamma$ by
fitting to the latest invariant mass spectrum from BESIII~\cite{Ablikim:2017fll}.
The amplitude is constructed according to Watson's
theorem, and the isospin-violating form factor is calculated within the framework of
R$\chi$T at LO in the $1/N_C$ expansion. We find that the
anomaly, defined in Eq.~(\ref{eq:anomaly}), is around $5.61\pm0.29$~GeV$^{-3}$.
It is shifted by an amount of $\mathcal{O}(1/N_C)$ compared to the value calculated
using the tree-level amplitude of WZW term. 
The couplings and decay widths
of the $\rho,\omega$ resonances are extracted properly, as shown in
Table~\ref{tab:pole}. The contributions to $\Gamma(\eta'\to \pi^+\pi^-\gamma)$
are quantified as follows:
$\Gamma(\eta'\to\rho\gamma\to \pi^+\pi^-\gamma)=56.6\pm5.3$~keV,
$\Gamma(\eta'\to\omega\gamma\to \pi^+\pi^-\gamma)=67.5\pm16.0$~eV,
and $\Gamma_{A}(\eta'\to \pi^+\pi^-\gamma)=3.34\pm 0.35$~keV.
The $\rho$ resonance dominates the intermediate process, while the anomaly
contributes more than the $\omega$. We obtain
$\Gamma(\eta'\to \rho\gamma)=56.6\pm5.3$~keV, $\Gamma(\eta'\to
\omega\gamma)=4.41\pm1.04$~keV, which are consistent with the determinations from
LO R$\chi$T within the uncertainty of around $1/3$.
Finally we find the phase of $P$-wave $\pi^+\pi^-$ scattering amplitude should have a
bump around $\sqrt{s}=M_\omega$. This work could be useful for
the studies of strong interaction referring to the $\pi\pi$ final
states, such as $J/\psi\to\gamma\pi\pi$~\cite{BES2006}, $\bar p p
\to X(3872) \to J/\psi\pi\pi$ in PANDA~\cite{PANDA2014}, and
$B_c^+\to B_s^0\pi^+\pi^0$ \cite{LXH2017}.

\section*{Acknowledgment}
We are grateful to Zhi-Hui Guo for helpful discussions.
We also wish to thank Chu-Wen Xiao for a careful reading of our manuscript and for his suggestions.
This work is supported in part by the Deutsche Forschungsgemeinschaft (DFG) and
the National Natural Science Foundation of China (NSFC) through
funds provided to the Sino-German CRC 110 ``Symmetries and the Emergence of Structure
in QCD''  (Grant No. TRR 110).
The work of UGM was supported in part by The Chinese Academy
of Sciences (CAS) President's International Fellowship Initiative (PIFI) with
Grant No. 2017VMA0025.
DLY thanks support from the Spanish Ministerio de Econom\'ia y Competitividad and
the European Regional Development Fund, under contracts FIS2014-51948-C2-1-P,
FIS2014-51948-C2-2-P, SEV-2014-0398 and by Generalitat Valenciana under
contract PROMETEOII/2014/0068.

$\,$

\appendix
\setcounter{equation}{0}
\setcounter{table}{0}
\renewcommand{\theequation}{\Alph{section}.\arabic{equation}}
\renewcommand{\thetable}{\Alph{section}.\arabic{table}}

\section{Isospin-violating form factors}\label{app:isov}
The amplitude of $\eta'\to\pi^+\pi^-\gamma$ is given by
\bea
\mathcal{M}_\lambda=e\epsilon^{\mu\nu\alpha\beta}\epsilon_\mu(k,\lambda) q_{\nu}p^+_\alpha p^-_\beta  F(s)\,.
\eea
For the form factor $F(s)$, the tree-level WZW anomaly term gives
\be
F_a=-\frac{N_C}{12 \sqrt{3}\pi^2 F^3}(\sin\theta_P+\sqrt{2}\cos\theta_P)\;. \label{eq:WZW}
\ee
The isospin-violating form factor is
\bea
F^{i.v.}_{tree}=F_b+F_c+F_d\,, \label{eq:A;Fiv}
\eea
and
\begin{widetext}
\bea
F_b&=&-\frac{8\sqrt{6}F_V\left(1+8\sqrt{2}\alpha_V\frac{m_{\pi}^2}{M_V^2}\right)}{3M_V F^3}\left(-\sin\delta+\frac{1}{\sqrt{3}}\cos\delta\sin\theta_V\right)\sin\delta(\sin\theta_P+\sqrt{2}\cos\theta_P) \no\\
&&G_{R\eta'}(0,s)BW_R[\omega,0] \, ,
\\[4mm]
F_c&=&\frac{4\sqrt{2}G_V}{3M_V F^3} \sin\delta\left\{\cos\delta\sin\theta_V [\sqrt{2}\cos\theta_P- \sin\theta_P ]+\sqrt{2}\cos\delta\cos\theta_V \sin\theta_P\right. \no\\
&&\left.-\sqrt{3} \sin\delta[\sqrt{2}\cos\theta_P+ \sin\theta_P]\right\}
C_{R\eta'1}(0,s,M_{\eta'}^2)BW_R[\omega,s] \no\\
&&+\frac{2\sqrt{2}G_V}{18 M_V F^3} \sin\delta
\left\{4 \cos\delta \left[-3\cos(\theta_V-\theta_P)+ \cos(\theta_V+\theta_P)+2\sqrt{2}\sin(\theta_V+\theta_P)\right] m_K^2 \right.\no\\
&&+\left[-6 \sqrt{3} \sin\delta [\sqrt{2}\cos\theta_P+ \sin\theta_P]+\cos\delta [ 9\cos(\theta_V-\theta_P)- \cos(\theta_V+\theta_P)\right.\no\\
&&\left.\left.-2\sqrt{2}\sin(\theta_V+\theta_P)]\right] M_\pi^2\right\} C_{R\eta'2} BW_R[\omega,s]\, ,
\\[4mm]
F_d&=&-\frac{8F_V\left(1+8\sqrt{2}\alpha_V\frac{M_\pi^2}{M_V^2}\right)G_V}{\sqrt{6} F^3 }\sin\delta\left(-\sin\delta+\frac{1}{\sqrt{3}}\cos\delta\sin\theta_V\right)\no\\
&&\left(\sin^2\delta (2 \cos\theta_P+\sqrt{2} \sin\theta_P)+\cos^2\delta~[~2 \cos\theta_P+\sin\theta_V (4 \cos\theta_V-\sqrt{2} \sin\theta_V ) \sin\theta_P~]\right)\no\\
&&D_{R\eta'1}(0,s,M_{\eta'}^2)BW_{RR}[\omega,\omega,0,s]\no\\
&&-\frac{4F_V\left(1+8\sqrt{2}\alpha_V\frac{M_\pi^2}{M_V^2}\right) G_V}{3\sqrt{6} F^3 }\sin\delta\left(-\sin\delta+\frac{1}{\sqrt{3}}\cos\delta\sin\theta_V\right)\no\\
&&\left\{4 \cos^2\delta \left(\cos\theta_P [3-\cos2\theta_V-2 \sqrt{2} \sin2\theta_V]+[3
\sqrt{2}-\sqrt{2} \cos2\theta_V-4 \sin2\theta_V] \sin\theta_P\right) m_K^2\right.\no\\
&&+\left(6 \sin^2\delta (2 \cos\theta_P+\sqrt{2} \sin\theta_P)+\cos^2\delta \left(4 \cos(2\theta_V+\theta_P)\right.\right.\no\\
&&\left.\left.+\sqrt{2} [8 \cos\theta_P \sin2\theta_V +(9-\cos2\theta_V) \sin\theta_P]\right)
m_{\pi }^2\right\}D_{R\eta'2}BW_{RR}[\omega,\omega,0,s]\no\\
&&-\frac{2F_V\left(1+8\sqrt{2}\alpha_V\frac{M_\pi^2}{M_V^2}\right) G_V}{\sqrt{6} F^3 }\sin\delta\left(\cos\delta+\frac{1}{\sqrt{3}}\sin\delta\sin\theta_V\right)\no\\
&&\left\{\sin2\delta [~-3\sqrt{2}+\sqrt{2} \cos2\theta_V+4 \sin2\theta_V~] \sin\theta_P\right\}
D_{R\eta' 1}(0,s,M_{\eta'}^2)BW_{RR}[\rho,\omega,0,s]\no\\
&&-\frac{2F_V\left(1+8\sqrt{2}\alpha_V\frac{M_\pi^2}{M_V^2}\right) G_V}{3\sqrt{6} F^3 }\sin\delta\left(\cos\delta+\frac{1}{\sqrt{3}}\sin\delta\sin\theta_V\right)
\sin2\delta\no\\
&&\left\{-4 \cos\theta_P \left(-3+\cos2\theta_V+2 \sqrt{2} \sin2\theta_V\right) \left(m_K^2-m_{\pi
}^2\right)\right.\no\\
&&\left.+\left(-3 \sqrt{2}+\sqrt{2} \cos2\theta_V+4 \sin[2\theta_V]\right) \sin\theta_P \left(4 m_K^2-m_{\pi
}^2\right)\right\}D_{R\eta' 2}BW_{RR}[\rho,\omega,0,s]\no\\
&&+\frac{2F_V\left(1+8\sqrt{2}\alpha_V\frac{2m_K^2-M_\pi^2}{M_V^2}\right) G_V}{3\sqrt{2} F^3 }\sin2\delta\no\\
&&\left\{\cos\theta_V \left(-4 \cos2\theta_V+\sqrt{2}
\sin2\theta_V\right) \sin\theta_P\right\}D_{R\eta' 1}(0,s,M_{\eta'}^2)BW_{RR}[\phi,\omega,0,s]\no\\
&&-\frac{2F_V\left(1+8\sqrt{2}\alpha_V\frac{2m_K^2-M_\pi^2}{M_V^2}\right) G_V}{9\sqrt{2} F^3 }\sin2\delta\cos\theta_V
\left\{4 \cos\theta_P ~[~2 \sqrt{2} \cos2\theta_V-\sin2\theta_V~]~(m_K^2-m_{\pi}^2)\right.\no\\
&&\left.-\left(4\cos2\theta_V-\sqrt{2} \sin2\theta_V\right) \sin\theta_P (4 m_K^2-m_{\pi}^2)\right\}D_{R\eta' 2}BW_{RR}[\phi,\omega,0,s]\;,
\eea
\end{widetext}
where
\begin{eqnarray}
&&C_{R\eta'1}(Q^2,x,m^2)=  (c_1-c_2+c_5) Q^2\no\\
&&\;\;\;\;\;\;\;\;\;\;\;\;-(c_1-c_2-c_5+2 c_6) x +(c_1+c_2-c_5)m^2 \; , \nonumber \\
&&C_{R\eta'2}= 8 \, c_3 \;  , \nonumber \\
&&D_{R\eta'1}(Q^2,x,m^2)=d_3(Q^2+x)+(d_1-d_3) \, m^2 \; , \nonumber \\
&&D_{R\eta'2}= 8 \, d_2 \; , \nonumber \\
&&G_{R\eta'}(Q^2,s)=(g_1+2g_2-g_3) \, (s-2m_{\pi }^2)\no\\
&&\;\;\;\;\;\;\;\;\;\;\;\;+g_2 \, (-Q^2+2m_{\pi }^2+m_{\eta }^2)+ (2g_4+g_5) \,  m_{\pi }^2 \; , \nonumber \\
&&BW_R[V,x]= \frac{1 }{M_{V }^2-i \Gamma_{V }(x) M_{V }-x} \, , \nonumber \\
&&BW_{RR}[V_1,V_2,x,y]= BW_R[V_1,x] \, BW_R[V_2,y] \, , \label{eq:BW}
\end{eqnarray}
Notice that $\Gamma_{V}(0)=0$. Here $\theta_V$ is the $\omega-\phi$ mixing angle, $\delta$ the $\rho-\omega$ mixing angle
and $\theta_P$ the $\eta-\eta'$ mixing angle.
We follow the construction of vector resonance off-shell widths in Ref.~\cite{GomezDumm:2000fz}, where the parametrization of $\Gamma_{\rho}(q^2)$ is employed as
\bea \label{eq:wrho}
\Gamma_\rho(q^2)&=&\frac{M_\rho \,  q^2}{96\pi F^2}\left[\left(1-\frac{4M_\pi^2}{q^2}\right)^{\frac{3}{2}} \theta(q^2-4M_\pi^2)\right. \no\\
&&\;\;\left.+\frac{1}{2}\left(1-\frac{4m_K^2}{q^2}\right)^{\frac{3}{2}}\theta(q^2-4m_K^2)\right] \,,
\eea
with $F$ the pion decay constant. For $\omega$, $\phi$ width we use constant decay widths.
The resonance parameters are given by PDG \cite{PDG2016}, and all other parameters can be found in Fit 4 of Ref.~\cite{DLY2013}.
For convenience we compile them in Table \ref{tab:2}. The form factors of $F_{\eta'\to\rho\gamma}$ and $F_{\eta'\to\omega\gamma}$ can be found in~\cite{DLY2013}, too.
\begin{table}[phtb!]
\begin{center}
\begin{tabular}{c c c c c}
\hline\hline
$F_V \, \mbox{(GeV)}$                 & 0.148$\pm$0.001       &   $\alpha_V$         &  0.0126$\pm$0.0007      \\
$2g_4+g_5$                            &  -0.493$\pm$0.003      &  $\theta_V(^\circ)$       &  38.94 $\pm$0.02        \\
$d_2$                                 &   0.0359$\pm$0.0007    &   $\theta_P(^\circ)$     &  -21.37$\pm$0.26     \\
$c_3$                                 &   0.00689$\pm$0.00017   &   $\delta(^\circ)$     &  2.12$\pm$0.06   \\
\hline
\end{tabular}
\caption{\label{tab:2} Parameters for $\rho-\omega$ mixing amplitudes.}
\end{center}
\end{table}
From the matching between R$\chi$T and QCD one can find the constraints on the unknown couplings \cite{RuizFemenia:2003hm,Dumm:2009kj}:
\begin{eqnarray}
&F_V G_V \,=\, F^2 \,, \nonumber \\
&d_1 + 8 d_2 -d_3 \,=\, \frac{F^2}{8 \, F_V^2}  \, , \nonumber \\
&4 c_3 + c_1 \,=\, 0 \, ,\nonumber \\
&d_3 \,=\, - \frac{N_C}{192 \, \pi^2} \frac{M_V^2}{F_V G_V} \, ,\nonumber \\
&c_1-c_2+c_5 \,=\, 0 \, ,\nonumber \\
 &g_2 \,=\, \frac{N_C}{192 \, \sqrt{2} \, \pi^2} \frac{M_V}{F_V} \, , \nonumber \\
&c_1+c_2+8c_3-c_5 \,=\, 0 \, ,\nonumber \\
&g_1 - g_3 \,=\, - \frac{N_C}{96 \, \sqrt{2} \, \pi^2} \frac{M_V}{F_V} \, ,\nonumber\\
&c_1-c_2-c_5+2c_6 \,=\, - \frac{N_C}{96 \sqrt{2} \pi^2} \frac{M_V}{G_V} \, . \label{eq:rescons5}
\end{eqnarray}
We notice that these constraints are not the same as a more complicated one \cite{Kampf:2011ty}, where heavier pseudoscalar resonances are  included. Since we only focus on the physics below $\sqrt{s}=M_{\eta'}$, we do not take the heavier resonances into consideration.
It would be interesting to note that our $F_V$ is closer to $F_V=\sqrt{3}F$ \cite{Guo:2010dv,Roig:2013baa} rather than  $F_V=\sqrt{2}F$ \cite{Ecker:1989yg}.
The former constraint is from the combined analysis of axial-vector current with the contribution from three pseudoscalars ($\tau\to KK\pi\nu_{\tau}$) and  $\tau\to \gamma P\nu_{\tau}$ (or $\langle VPP \rangle$ ), while our constraints are based on the analysis of vector current with the contribution from three pseudoscalars ($e^+e^-\to\pi^+\pi^-\pi^0/\eta$).
A more careful study is needed for a good understanding of the short-distance QCD constraint.




\end{document}